\documentclass[%
 aip, apl,
 amsmath,amssymb,
 reprint,%
]{revtex4-1}

\usepackage{graphicx}
\usepackage{dcolumn}
\usepackage{bm}

\begin{document}
\preprint{AIP/123-QED}
\title{Effect of nanostructure on thermoelectric properties of La$_{0.7}$Sr$_{0.3}$MnO$_{3}$ in 300-600 K range} 
\author{Saurabh Singh}
\email[Electronic mail: ]{saurabhsingh950@gmail.com}
\affiliation{School of Engineering, Indian Institute of Technology Mandi, Kamand - 175005, India}
\author{Simant Kumar Srivastav}
\affiliation{Department of Physics, Indian Institute of Technology Delhi, Hauz Khas, New Delhi-110016, India}
\author{Ashutosh Patel}
\affiliation{School of Engineering, Indian Institute of Technology Mandi, Kamand - 175005, India}
\author{Ratnamala Chatterjee}
\affiliation{Department of Physics, Indian Institute of Technology Delhi, Hauz Khas, New Delhi-110016, India}
\author{Sudhir Kumar Pandey}
\affiliation{School of Engineering, Indian Institute of Technology Mandi, Kamand - 175005, India}

\date{\today}

\begin{abstract}
In oxide materials, nanostructuring effect has been found very promising approach for the enhancement of \textit{figure-of-merit}, \textit{ZT}. In the present work, we have synthesized La$_{0.7}$Sr$_{0.3}$MnO$_{3}$ (LSMO) compound using sol-gel method and samples of crystallite size of $\sim$20, $\sim$41, and $\sim$49 nm were obtained by giving different heat treatment. Seebeck coefficient ($\alpha$), electrical resistivity ($\rho$), and thermal conductivity ($\kappa$) measurements were carried out in 300-600 K temperature range. The systematic change in the values of $\alpha$ from $\sim$ -19 $\mu$V/K to $\sim$ -24 $\mu$V/K and drastic reduction in the values of $\kappa$ from $\sim$0.88 W/mK to $\sim$0.23 W/mK are observed as crystallite size is reduced from 49 nm to 20 nm at $\sim$600 K. Also, fall in the values of $\rho$ in the paramagnetic (PM) insulator phase (400-600 K) are effectively responsible for the increasing trend in the values of \textit{ZT} at high temperature. For the crystallite size of 41 nm, value of \textit{ZT} at 600 K was found to be $\sim$0.017, which can be further increased up to $\sim$0.045 around 650 K temperature. The predicted value of \textit{ZT} suggests that LSMO can be suitable oxide material for thermoelectric applications at high temperature.
\end{abstract}


\maketitle 
In the past few decades, thermoelectric (TE) materials have been investigated extensively for an alternate and renewable source of energy.\cite{disalvo, bell} In the search of new materials, oxide materials have attracted much attention in the field due to their non-toxicity, oxidation resistance, high-temperature stability, easy and low cost manufacturing factors.\cite{koumoto} A material is said to be suitable for the TE application on the basis of \textit{figure of merit}, \textit{ZT}, which is defined as \textit{ZT} = ($\alpha$$^{2}$$\sigma$/$\kappa$), where the terms $\alpha$, $\sigma$, and $\kappa$ are Seebeck coefficient (or thermopower), electrical conductivity (inverse of electrical resistivity, $\rho$), and thermal conductivity, respectively.\cite{pei, lalonde} There are two different source of contributions in the total $\kappa$, defined as $\kappa$ = $\kappa$$_{e}$ + $\kappa$$_{l}$, where $\kappa$$_{e}$ and $\kappa$$_{l}$ are known as electronic and lattice thermal conductivity, respectively. The expression of \textit{ZT} suggests that, the magnitude of $\alpha$ and $\sigma$ should be larger; whereas, lower value of $\kappa$ (especially $\kappa$$_{l}$) is required for the higher values of \textit{ZT}.\cite{nolas} In search of materials with high \textit{ZT} value, many experimental and theoretical approaches have been used, few of them are such as making the materials with appropriate combination of elements, suitable doping, lowering the dimension, creating defects mechanism, nano structuring and band engineering, etc.\cite{minnich} Among these, nano structuring method has been one of the novel and effective approach for getting the higher \textit{ZT} values; as decreasing the grain size to the nano scale region increases the phonon scattering in the intragranular region. Due to this scattering effect, phonon mean free path reduces and results in to the decrement in the value of $\kappa$$_{l}$.\cite{dong} In many nanocrystalline size materials, over all values of $\kappa$ were found to be much lower than that of corresponding bulk or single crystal material.\cite{nan}\\
Generally, oxide materials have limitations for the TE applications due to its large value of $\kappa$. From the industrial point of view, the mechanism by which values of $\kappa$ can be reduced in the oxide materials, with an optimized value of power factor ($\alpha$$^{2}$$\sigma$), are highly demanded and play an important role in tuning the material properties for the TE applications. The size effect on TE properties have been seen in the oxide materials where lowering the crstallite size in nm range increases the magnitude of $\alpha$, whereas drastically reduce the values of $\kappa$.\cite{dura} The aspects of size effect on TE properties of La$_{0.7}$Sr$_{0.3}$MnO$_{3}$ (LSMO) samples have been found to be of interest due to the improvement in the magnitude of $\alpha$ with reduction in crystallite size.\cite{salazar} The study of TE properties of this compound by Salzar \textit{et al}. were done on the smallest possible crystallite size of 73 nm, and the measurements of $\alpha$ and $\sigma$ were reported only upto 500 K. There are various report on the TE measurement of bulk sample in wide temperature range.\cite{mahendiran, ohtani} In high temperature region, 300-400 K, the measurement on micro-crystallite size sample shows a large value of $\kappa$ ($\sim$2.4 W/mK, at 300 K), and limits its use for TE application.\cite{wang} In this situation, the possible solution can be the reduction of $\kappa$ by lowering the lattice thermal conductivity and improving the magnitude of $\alpha$ while having the moderate value of $\rho$. Also, the effect of nano crystallite size on the value of \textit{ZT} were found to be lacking for this compound in the literature for the high temperature region, to the best of our knowledge. With this motivation, TE study on nano-crystallite size LSMO samples were investigated in the high temperature region.\\ 
In the present work, La$_{0.7}$Sr$_{0.3}$MnO$_{3}$ compound was synthesized using sol-gel method and samples with three different crystallite size in nano scale range (i.e. 20, 41, and 49 nm) were obtained by sintering the samples at different temperature and time. Measurements of $\alpha$, $\rho$, and $\kappa$ were carried out in the 300-600 K temperature range and values of \textit{ZT} were also estimated to see the applicability of this material for TE applications.\\ 
 Nanoparticles of La$_{0.7}$Sr$_{0.3}$MnO$_{3}$ (LSMO) were prepared by the sol–gel method.\cite{srivastav} Lanthanum nitrate hexahydrate, La(NO$_{3}$)$_{3}$.6H$_{2}$O, manganese nitrate tetrahydrate, Mn(NO$_{3}$)$_{2}$.4H$_{2}$O, strontium nitrate, Sr(NO$_{3}$)$_{2}$ (all from Sigma-Aldrich, USA), propylene glycol (Thomas Baker, India) and citric acid (Merck, Germany) were of analytical grade. Stoichiometric amount of metal nitrates were dissolved in deionized water separately and then mixed to prepare precursor solution. Further, propylene glycol and citric acid were added to the above precursor solution in 1:1 mole ratio with respect to metal nitrates. The solution was magnetically stirred and heated on a thermal plate at $\sim$90 $^{o}$C until all the liquid evaporated out and black precursor powder was obtained. In order to get the well crystallized LSMO nanoparticles, the precursor powder was calcined at 800 $^{o}$C for 3 hrs in ambient atmosphere with heating rate of 2 $^{o}$C min$^{-1}$. To obtain the samples with different grain size, powder form of samples were pelletized in 5 mm diameter pellet under the pressure of $\sim$40 kg/cm$^{2}$, and pellets were sintered at 800 $^{o}$C (24 hr) and 850 $^{o}$C (72 hr). For the structural characterization, XRD diffraction pattern were recorded using the Rigaku Advance x-ray diffractometer using Cu K$\alpha$ radiation ($\lambda$= 1.5418 \AA). The temperature dependent measurement of $\alpha$, $\rho$, and $\kappa$ were carried out using the home made setup.\cite{Patel, resistivitysingh}\\
Fig. 1 shows the XRD pattern of all the samples. To analyze the x-ray data, Rietveld refinement were performed using the FULLPROF software.\cite{rodriguez} From the refinement, goodness of fit $\chi$$^{2}$ was achieved to the 1.25, and the values of lattice parameters of the unit cell were \textit{a} = 5.488(3) $\AA$ and \textit{c} = 13.371(5) $\AA$. The refinement results (shown in the inset of Fig. 1) corresponding to rhombohedral structure described by R-3C space group confirms that sample is of single phase.\cite{radaelli} 
\begin{figure}
\includegraphics[width=0.45\textwidth]{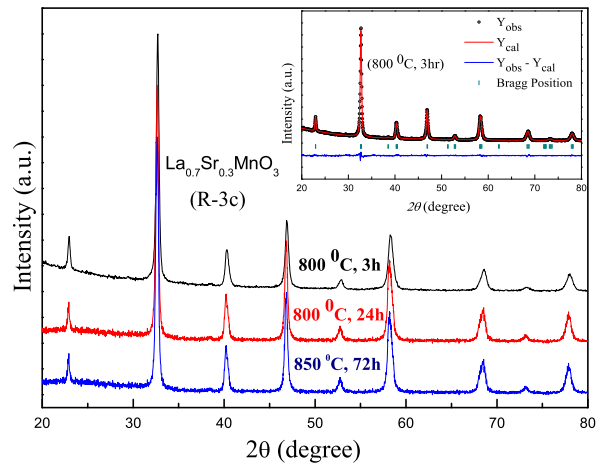}%
\caption{(Color online)XRD pattern of La$_{0.7}$Sr$_{0.3}$MnO$_{3}$ sample. Rietveld refinement of XRD data (800 $^{o}$C, 3h) is shown in the inset of the figure.}%
\end{figure}
The values of crystallite size (D) were calculated from the half-width (FWHM- Full Width Half Maximum) of most intense peak, using the Debye Scherrer formula, D = [(k$\lambda$)/(B(\textit{2$\theta$}).Cos $\theta$)], where k is a constant (0.94), B(\textit{2$\theta$}) is FWHM, and $\lambda$ is 1.5418 \AA.\cite{warren} The estimated values of crystallite size were found to be 20 nm, 41 nm, and 49 nm, for the sample sintered at 800 $^{o}$C (3hr), 800 $^{o}$C (24 hr), and 850 $^{o}$C (72 hr), respectively. We have also estimated the particle size of sample (LSMO 800 $^{o}$C, 3hr) from the TEM  (transmission electron microscope) image analysis (see supplementary material), and the value of mean particle size was found to be $\sim$20 nm, which is consistent with the value obtained from its XRD result.\cite{kumar}\\
Temperature dependent variations of the measured values of $\alpha$ in 300-600 K are shown in Fig. 2.
\begin{figure}
 \vspace{1.2cm}
\includegraphics[width=0.35\textwidth]{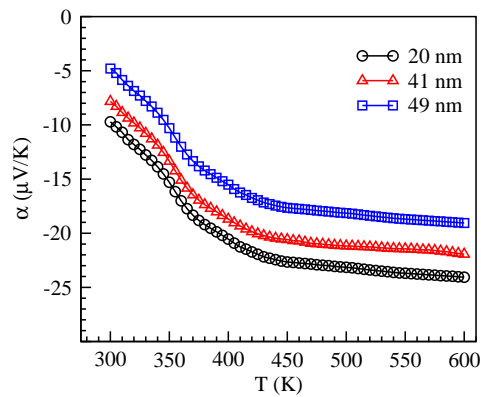}%
\caption{(Color online) Seebeck coefficient, $\alpha$(T), variation with temperature.}%
\end{figure} 
In 300-450 K, an increasing trends in the magnitude of $\alpha$ were noticed for all the samples, where as above 450 K, the values of $\alpha$ were almost constant up to 600 K. The negative values of $\alpha$ for all the samples were observed in the entire temperature range under study, which suggests that electrons are dominant carriers in the contribution of $\alpha$ and this system have the character of \textit{n-type} TE material. At 600 K, the values of $\alpha$ are $\sim$-19, $\sim$-22, and $\sim$ -24 $\mu$V/K for the 20 nm, 41 nm, and 49 nm samples, respectively. These values are larger than the reported value of $\alpha$ ($\sim$ -17 $\mu$V/K at $\sim$470 K) for 73 nm crystallite size sample.\cite{salazar} \textit{ZT} of the materials are dependent on square of the $\alpha$ values, therefore observation of increment in the magnitude of $\alpha$ with decrease in the crystallite size of the LSMO suggests that this compound can be good TE material.\\
  For the different crystallite size samples, electrical resistivity ($\rho$) vs. temperature plot are shown in the Fig. 3.
\begin{figure}
\includegraphics[width=0.35\textwidth]{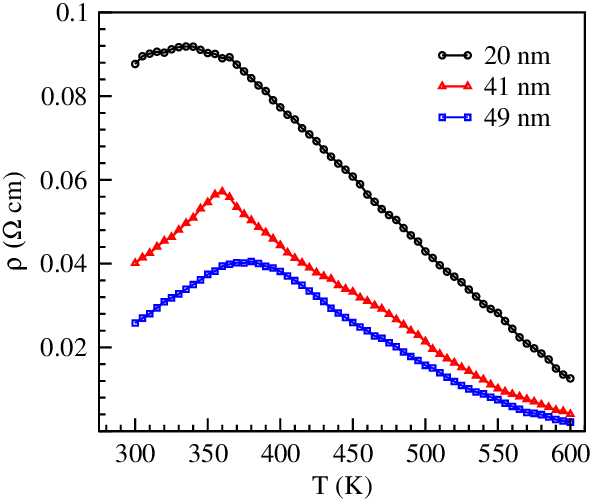}%
\caption{(Color online) Resistivity, $\rho$(T), vs. T behavior of samples with grain size 20 nm, 41 nm, and 49 nm.}%
\end{figure}
From the Fig. 3, it is observed that with decrease in crystallite size from 49 nm to 20 nm, the magnitude of resistivity increases relatively, whereas metal to insulator transition temperature (T$_{MI}$) shift towards the lower temperature. This type of behavior have also been noticed in the LSMO and other oxide materials.\cite{dey, salazar} The values of T$_{MI}$ are at $\sim$340 K, $\sim$358 K, and $\sim$370 K for the 20 nm, 41 nm, and 49 nm samples, respectively. The values of $\rho$ are found to be in same order of the reported values of bulk sample ($\rho$ = $\sim$0.08 $\Omega$ cm at 300 K).\cite{taran} In the insulating region, 400-600 K, the values of $\rho$ decreases with temperature and there are small difference between the magnitude of $\rho$ of 41 nm and 49 nm samples were observed. The overall behavior of $\rho$ and variations in its magnitude from 20 nm to 49 nm samples suggest that this compound has good electrical property and can be useful for the TE applications.\\
Fig. 4 shows the temperature dependent behavior of total thermal conductivity ($\kappa$) plot of the 20 nm, 41 nm, and 49 nm samples in the 300-600 K temperature range.
\begin{figure}
 \vspace{1.5cm}
\includegraphics[width=0.35\textwidth]{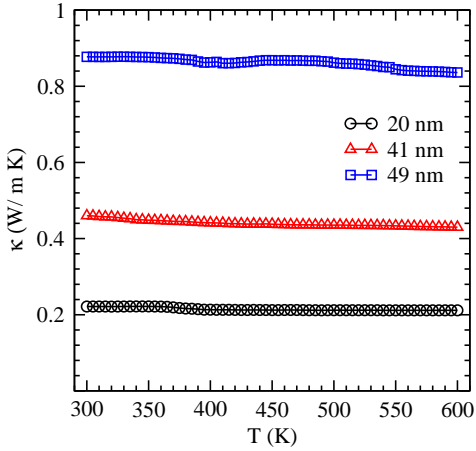}%
\caption{(Color online) Temperature dependent variation of thermal conductivity of La$_{0.7}$Sr$_{0.3}$MnO$_{3}$ with 20 nm, 41 nm, and 49 nm crystallite size samples.}%
\end{figure}
For the different grain size samples, $\kappa$ have very weak dependent on temperature and found to be almost linear in the whole temperature range under study. At 300 K, the observed values of $\kappa$ are $\sim$0.89, $\sim$0.45, and $\sim$0.23 W/mK for the 49 nm, 41 nm, and 20 nm samples, respectively. With decrease in grain size from 49 nm to 20 nm, the systematic decrements in the values of $\kappa$ are found. These values of thermal conductivity are very small in comparison to the value of $\kappa$ reported for bulk sample ($\kappa$ = $\sim$2.4 W/m K at 300 K). The drastic decrements in $\kappa$ values are due to the increase in phonon scattering with decrease in crystallite size, which lower the contributions of lattice part to the total thermal conductivity. The values of $\kappa$ for the 49 nm, 41 nm, and 20 nm samples are $\sim$63\%, $\sim$81\%, and $\sim$90\% smaller than that of the reported values of microcrystalline samples. This kind of behavior have been also seen in the similar oxide material.\cite{dura} For La$_{0.7}$Sr$_{0.3}$Mn$O_{3}$, the huge reduction in the $\kappa$ values for the samples with crystallite size of nm range is reported first time. Minimization of $\kappa$ value is one of the challenging task and also is the most crucial requirement to get the higher value of \textit{ZT}. Synthesizing the sample with crystallite size of 20 nm, the value of $\kappa$ of LSMO has been reduced to one-tenth of its value reported at 300 K, which is a very good signature for TE applications.\\ 
In order to know the potential capabilities of La$_{0.7}$Sr$_{0.3}$MnO$_{3}$ for TE applications, we have also estimated the \textit{ZT} values in the 300-600 K temperature range, which are shown in the Fig. 5.
\begin{figure}
\includegraphics[width=0.40\textwidth]{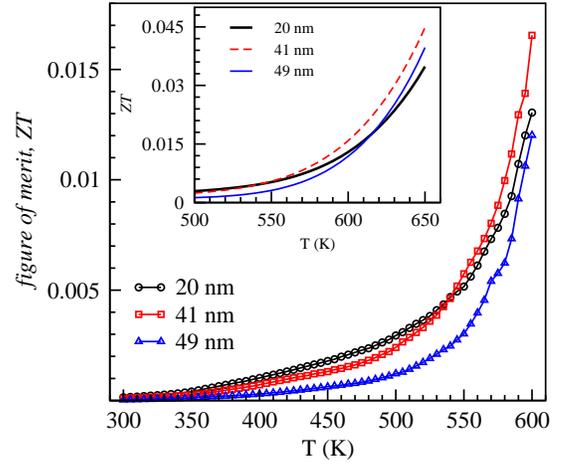}%
\caption{(Color online)Temperature variation of \textit{figure of merit}, \textit{ZT}. Inset shows the predicted data up to 650 K}%
\end{figure}
The values of \textit{ZT} are found to be increasing monotonically with temperature in 300-500 K range, and in this temperature range the values of \textit{ZT} of 20 nm sample are found to be larger than that of 41 nm and 49 nm sample. Above 500 K temperature, \textit{ZT} values of all three samples increases sharply up to 600 K due to almost constant value of $\alpha$ and $\kappa$, while having the continuous decrements in the $\rho$(T) values. The \textit{ZT} curve of 41 nm sample make crossover of the 20 nm sample at $\sim$540 K and reaches to the maximum value of \textit{ZT} equal to $\sim$0.017 at $\sim$600 K. The optimized values of $\alpha$, $\rho$ and $\kappa$ of the 41 nm sample is responsible for showing the higher values of ZT in 540-600 K range. At 600 K, sample with 41 nm grain size have the highest \textit{ZT} value than that of 20 nm (\textit{ZT} = $\sim$0.013) and 49 nm (\textit{ZT} = $\sim$0.012) samples.\\
In the high-temperature region of the PM insulator phase, with almost constant values of $\alpha$ and $\kappa$, the continuous decrease in the values of $\rho$ make it a key parameter for increasing the values of \textit{ZT} in LSMO compound. In PMI region, any electronic phase transition is not reported for LSMO below 650 K. Thus, we can also expect the similar temperature dependent behavior of $\alpha$, $\rho$, and $\kappa$ above 600 K. Considering the similar trend of $\alpha$, $\rho$, and $\kappa$, the values of \textit{ZT} have been estimated up to 650 K, which is shown in the inset of Fig. 5. The predicted values of \textit{ZT} for 20 nm, 41 nm, and 49 nm samples are 0.035, 0.045, and 0.040, respectively at 650 K, which are nearly three times of the observed values at 600 K. It will be more interesting to see the experimental verification of predicted value of \textit{ZT} at 650 K. To confirm this conjecture, further measurement of $\alpha$, $\rho$ and $\kappa$ on nano-crystallite LSMO samples above 600 K are highly desirable.
\\
In conclusion, we have prepared the La$_{0.7}$sr$_{0.3}$MnO$_{3}$ using the sol-gel method, and three different samples of grain sizes 20 nm, 41 nm, and 49 nm were obtained by giving different heat treatment. For this system, lowering the grain size of samples in nano-meter range were found to be very effective in minimization of thermal conductivity, increment in magnitude of Seebeck coefficient, whereas decreasing trends in magnitude of resistivity in the paramagnetic insulator region is noticed.Particularly, for 41 nm sample, values of \textit{ZT} were reported to $\sim$0.017 at 600 K, and also predicted to be improved up to $\sim$0.045 at 650 K.  For the 41 nm sample, the continuous increase in the values of \textit{ZT} with temperature and its reported value suggest that La$_{0.7}$Sr$_{0.3}$MnO$_{3}$ compound can be used for the TE applications in the high temperature region. The present study shows that nanostructuring approach is very novel and found to be very effective for the improvement of TE properties of oxide materials. 
\bibliography{aipsamp}
\end{document}